\documentclass[11pt]{article}

\usepackage{amsmath}
\usepackage{amsfonts}
\usepackage{epsfig}
\usepackage{latexsym}

\newcommand{\be}{\begin{equation}}
\newcommand{\ee}{\end{equation}}
\newcommand{\bea}{\begin{eqnarray}}
\newcommand{\eea}{\end{eqnarray}}

\begin{document}

\bibliographystyle{hieeetr}

\begin{titlepage}
\begin{flushright}
hep-th/0501054
\end{flushright}
\vskip 1in
\begin{center}
{\Large{Falling D0-Branes in 2D Superstring Theory}} \vskip 0.5in
{Joshua M. Lapan, Wei Li}
\vskip 0.3in {\it Jefferson Physical Laboratory,
Harvard University\\
Cambridge, MA  02138
}
\end{center}
\vskip 0.5in

\begin{abstract}

In $\mathcal{N}=1$, 2D superstring theory in the linear dilaton background, there exists falling D0-branes that are described by time-dependent boundary states.  These falling D0-brane boundary states can be obtained by adapting the FZZT boundary states of $\mathcal{N}=2$ Super Liouville Field Theory (SLFT) \cite{Nakayama:2004yx} to the case of the $\mathcal{N}=1$, 2D superstring.  In particular, we find that there are four stable, falling D0-branes (two branes and two anti-branes) in the Type 0A projection and two unstable ones in the Type 0B projection, leaving us with a puzzle for the matrix model dual of the theory.

\end{abstract}

\end{titlepage}

\tableofcontents

\section{Introduction}

In both bosonic and supersymmetric Liouville Field Theory (LFT),
there exist static D0 and D1-branes.  In particular, the static D0-branes---the so-called ZZ branes---sit in the strong coupling region $\phi\rightarrow +\infty$ \cite{Fateev:2000ik}.  In the bosonic system with Euclidean time, Lukyanov, Vitchev, and Zamolodchikov, showed the existence of a time-dependent boundary state, the paperclip brane, that breaks into two hairpin-shaped branes in the UV region \cite{Lukyanov:2003nj}.  They derived the wave function of the boundary state from the classical shape of the brane in the spacetime.  Under the Wick-rotation from Euclidean time into Minkowski time, the hairpin brane is reinterpreted as the falling D0-brane.

The falling D0-brane in the supersymmetric system was first considered by Kutasov \cite{Kutasov:2004dj}.  He studied the classical dynamics of the falling D0-brane in the vicinity of a stack of NS5-branes that produce a linear dilaton background.  In his treatment, the radial position (along the Liouville direction) of the D0-brane is a dynamic field living on its worldvolume, and so the corresponding DBI action gives the classical trajectory of the D-brane in this background:
\be
\label{eq:classical trajectory}
e^{-\frac{Q\phi}{2}}=\frac{\tau_p}{E}\cosh\frac{Qt}{2}~,
\ee
where $Q$ is the background charge of the linear dilaton, and $\tau_p$ and $E$ are the tension and energy of the D0-brane, respectively.

In $\mathcal{N}=2$ SLFT, which has Euclidean time, a type of time-dependent boundary state solution was derived in \cite{Eguchi:2003ik}.  The Wick-rotation of the above solution was carried out in \cite{Nakayama:2004yx} to study the wave function of the falling D0-brane in the $\mathcal{N}=2$ SLFT system.  The wave function reproduced the trajectory (\ref{eq:classical trajectory}) in the classical limit, leading to the suggestion that this was the supersymmeterized version of the hairpin brane.

We show that in $\mathcal{N}=1$, 2D superstring theory with a linear dilaton background---which we will use interchangeably with $\hat{c}_m=1~\mathcal{N}=1$ SLFT---there exists a similar, time-dependent boundary state corresponding to the falling D0-brane.  The naive argument for the existence of the falling D0-brane is as follows.  As is well known, the mass of the D0-brane is inversely related to the string coupling as
\be
\label{eq:D-brane mass}
m=\frac {1}{g_s} = e^{-\phi}~,
\ee
so the mass of the D0-brane decreases as it runs along the Liouville direction from the weak coupling region ($\phi \rightarrow -\infty$) to the strong coupling region ($\phi \rightarrow +\infty$).  Thus, if we set a D0-brane free at the weak coupling region, it will roll along the Liouville direction towards the strong coupling region until it reaches the Liouville wall.  This is the falling D0-brane solution which can be described by a time-dependent closed string boundary state of the $\mathcal{N}=1$, 2D superstring.

In the bosonic case, the hairpin brane satisfies symmetries in addition to those of the action (conformal symmetry).  The additional symmetry is known as the $\mathcal{W}$-symmetry and is generated by higher spin currents \cite{Lukyanov:2003nj}.  The hairpin brane is then constructed from the integral equations that are defined by the $\mathcal{W}$-symmetry.  In the $\mathcal{N}=1$, 2D superstring, it should be possible to use the supersymmeterized version of the $\mathcal{W}$-symmetry to go through a similar construction and find a falling D0-brane.  However, we will argue that it can also be obtained by adapting the falling D0-brane solution in $\mathcal{N}=2$ SLFT \cite{Nakayama:2004yx}, \cite{Eguchi:2003ik}, to the $\mathcal{N}=1$, 2D superstring.

In section \ref{sec:N=1}, we briefly review properties of $\mathcal{N}=1$ SLFT, including the construction of boundary states corresponding to ZZ-branes and FZZT-branes using the modular bootstrap approach.  In section \ref{sec:N=2}, we review properties of $\mathcal{N}=2$ SLFT as well as the construction of the boundary state corresponding to the falling D0-brane.  Finally, in section \ref{sec:Falling D0-Brane} we argue that we may slightly modify the $\mathcal{N}=2$ SLFT falling D0-brane boundary state to obtain the solution in $\mathcal{N}=1$, 2D superstring theory, and we discuss the number of falling D0-branes in the Type 0A and 0B projections.  It would be interesting to understand these falling D0-branes in the context of matrix models, but this is beyond the scope of this paper.

\section{$\mathcal{N}=1$, 2D Superstring Theory and its Boundary States}
\label{sec:N=1}

\subsection{$\hat{c}_m=1~\mathcal{N}=1$ SLFT}
\label{subsec:c=1 N=1 SLFT}

The $\mathcal{N}=1$ super Liouville theory can be obtained by the quantization of a two dimensional supergravity theory \cite{Polyakov:1981re}.  After eliminating the auxiliary field by its equation of motion, adding $\hat{c}_m=1$ matter, and setting $\alpha'=2$, the free part of the action is
\be
\label{eq:N=1 action}
S_0 = \frac{1}{2\pi}\int d^2z \left[\delta_{\mu\nu}\left(\partial X^\mu\bar{\partial} X^\nu + \psi^\mu\bar{\partial}\psi^\nu + \tilde{\psi}^\mu\partial\tilde{\psi}^\nu \right) + \frac{Q}{4}RX^1 \right]~,
\ee
where $\mu,~\nu=1,~2$.  Since we are considering 2D superstring theory below, we will write $\phi=X^1$ and $Y=X^2$ as is common in the literature.  The $\mathcal{N}=1$ SLFT also includes a potential term
\be
S_{int}^{\mathcal{N}=1} = 2i\mu b^2 \int d^2z~ :e^{b\phi}: :\left(\psi^1\tilde{\psi}^1 + 2\pi\mu e^{b\phi}\right):~,
\ee
where we must have $Q=b+\frac{1}{b}$ for conformal invariance (note that the normal ordering is crucial for this result and comes from the elimination of the auxiliary field).

The stress energy tensor and superconformal current are
\bea
\label{eq:N=1 generators}
T &=& -\frac{1}{2} \partial Y \partial Y -\frac{1}{2} \partial \phi
\partial \phi +\frac{Q}{2} \partial^2 \phi -\frac{1}{2}\delta_{\mu\nu}\psi^\mu \partial \psi^\nu   \nonumber \\ 
G &=& i(\psi^1\partial \phi + \psi^2\partial Y - Q\partial \psi^1)~,
\eea
which produce the $\mathcal{N}=1$ superconformal algebra (SCA)
\bea
\label{eq:N=1 SCA algebra}
\left[ L_m , L_n \right] & = & (m-n) L_{m+n} + \frac{c}{12}(m^3-m)\delta_{m,-n} \nonumber \\
\{ G_r , G_s \} & = & 2L_{r+s} +\frac{c}{12}(4r^2-1)\delta_{r,-s} \nonumber \\
\left[ L_m , G_r \right] & = & \frac{m-2r}{2} G_{m+r}~,
\eea
where $c = \frac{3}{2} \hat{c} = \frac{3}{2}(1+2Q^2+1)$, and $r$ and $s$ take integer (half-integer) values in the R (NS) sector.  For a critical string theory, we must set $Q=2$, corresponding to $b=1$.

The primary fields in the NS sector are $N_{p,\omega } = :e^{(\frac{Q}{2} + ip) \phi + i \omega Y }:$ with weights $h_{p,\omega}^{\textrm{NS}}=\frac{1}{2}(\frac{Q^2}{4}+p^2+\omega^2)$, while in the R sector they are $R_{p,\omega}^{\pm} = \sigma ^{\pm}N_{p,\omega}$ with weights $h_{p,\omega}^{\textrm{R}}=h_{p,\omega }^{\textrm{NS}}+\frac{1}{16}$.  (If we bosonize the complex fermion $\psi^\pm=\frac{1}{\sqrt{2}}(\psi^2\pm i\psi^1)=:e^{\pm iH}:$, then $\sigma^\pm$ is given by $\sigma^\pm = :e^{\pm iH/2}:$.)  The open string character is
\be
\label{eq:N=1 character}
\chi_{p,\omega}^{\sigma,\pm}(\tau)=\mathrm{Tr}_{\mathcal{H}_{p,\omega}^\sigma}[q^{L_0-c/24}(\pm1)^F]~,
\ee
where $q\equiv e^{2\pi i\tau}$ and we trace over the descendants of $N_{p,\omega}$ or $R_{p,\omega}$ for $\sigma=\textrm{NS},~\textrm{R}$, respectively.  For non-degenerate representations, the open string characters (which result from a trace over a corresponding primary state and its descendants) are \cite{Ahn:2002ev}, \cite{Nakayama:2004vk},
\bea
\label{eq:N=1 characters}
\chi_{p,\omega}^{\textrm{NS},+}(\tau) &=&  q^{\frac{1}{2}(p^2+\omega^2)}
\frac{\theta_{00}(\tau,0)}{\eta(\tau)^3} \nonumber \\
\chi_{p,\omega}^{\textrm{NS},-}(\tau) &=&  q^{\frac{1}{2}(p^2+\omega^2)}
\frac{\theta_{01}(\tau,0)}{\eta(\tau)^3} \nonumber \\
\chi_{p,\omega}^{\textrm{R},+}(\tau) &=&  q^{\frac{1}{2}(p^2+\omega^2)}
\frac{\theta_{10}(\tau,0)}{\eta(\tau)^3} \nonumber \\
\chi_{p,\omega}^{\textrm{R},-}(\tau) &=& 0~.
\eea

\subsection{Open/Closed Duality: Boundary States}

As is well known, we can realize boundary conditions for an open string as constraints on states in the closed string spectrum \cite{DiVecchia:1999rh}, \cite{DiVecchia:1999fx}, \cite{Gaberdiel:2000jr}, and \cite{Nepomechie:2001bu}.  For example, Neumann and Dirichlet boundary conditions in the open string are realized classically as $\partial X(y) \mp \bar{\partial}X(y) = 0$ and $\psi(y) \mp \eta\tilde{\psi}(y) = 0$, where $\eta=\pm 1$, $y\in\mathbf{R}$, and we have taken the boundary to lie along the real axis (the upper sign is for Neumann boundary conditions and the lower for Dirichlet).  To tranform from the open channel to the closed channel, we must perform the coordinate transformation $z\rightarrow z$ and $\bar{z}\rightarrow \bar{z}^{-1}$, resulting in the conditions $\partial X(y) \pm y^{-2}\bar{\partial}X(y^{-1})=0$ and $\psi(y) \pm i\eta y^{-1}\tilde{\psi}(y^{-1})=0$.  When we quantize the theory, these become constraints on closed string boundary states:
\bea
\label{eq:ND bcs}
\textrm{Neumann:} & \qquad & (\alpha_m + \tilde{\alpha}_{-m})|B,\eta\rangle=(\psi_r + i\eta\tilde{\psi}_{-r})|B,\eta\rangle=0  \nonumber \\
\textrm{Dirichlet:} & \qquad & (\alpha_m - \tilde{\alpha}_{-m})|B,\eta\rangle=(\psi_r - i\eta\tilde{\psi}_{-r})|B,\eta\rangle=0~.
\eea

In general, for a state in the closed string Hilbert space to be a boundary state, it must satisfy two conditions.  First, the state must satisfy contraints coming from the requirement that the corresponding boundary vertex operator preserve the symmetries of the original theory.  In the case of a simple bosonic theory, this amounts to requiring conformal invariance which, in the language of boundary states, translates to the constraints
\be
(L_m - \tilde{L}_{-m})|B\rangle=0~.
\ee
Second, the state must satisfy constraints coming from the open/closed duality of a cylinder diagram.  If an open string satisfies some boundary conditions $\alpha$ and $\beta$ on its left and right ends, respectively, then the corresponding closed string boundary states must satisfy Cardy's condition \cite{Cardy:1989ir}:
\be
\label{eq:open-closed}
\langle B,\alpha|q_c^{\frac{1}{2}H_c}|B,\beta\rangle = \textrm{Tr}_{\mathcal{H}_{\alpha\beta}}[q_o^{H_o}]~,
\ee
where $H_c$ and $H_o$ are the closed and open string Hamiltonians, $q_c=e^{2\pi i\tau_c}$ and $q_o=e^{2\pi i\tau_o}$, and the trace on the right is taken over the open string spectrum that satisfies the specified boundary conditions, $\mathcal{H}_{\alpha\beta}$.  The open and closed string moduli are related through worldsheet duality by a modular transformation, $\tau_c=-\frac{1}{\tau_o}$.

\subsection{Ishibashi and Cardy States: The Modular Bootstrap}

In $\hat{c}_m=1~\mathcal{N}=1$ SLFT, a closed string boundary state must satisfy \cite{Ahn:2002ev}, \cite{Nakayama:2004vk},
\bea
\label{eq:N=1 b.s.}
(L_{m} - \tilde{L}_{-m}) |B,\alpha;\eta,\sigma \rangle &=& 0 \nonumber \\
(G_{r}-i\eta\tilde{G}_{-r}) |B,\alpha;\eta,\sigma \rangle &=& 0~,
\eea
where $\alpha$ labels an open string conformal family, $\sigma=\textrm{NS},~\textrm{R}$ (really NS-NS and R-R, but we will commonly use this short-hand), and $\eta=\pm$ gives the spin structure of the boundary states.  Additionally, the boundary states must satisfy the open/closed duality requirement (\ref{eq:open-closed}).  These states are commonly referred to as the Cardy states.

To find the Cardy states, it is convenient to use an orthonormal basis of states satisfying (\ref{eq:N=1 b.s.}).  The so-called Ishibashi states $|i;\eta,\sigma\rangle\rangle$ of the theory form such a basis \cite{Ishibashi:1988kg} and are defined to satisfy the additional constraints
\be
\label{eq:N=1 Ishibashi definition}
\langle \langle i;\eta,\sigma|q_c^{\frac{1}{2}H_c}|j;\eta',\sigma' \rangle\rangle = \delta_{ij}\delta_{\sigma\sigma'}\chi_i^{\sigma,\eta\eta'}(\tau_c) \equiv \delta_{ij}\delta_{\sigma\sigma'}\mathrm{Tr}_{\mathcal{H}_i^\sigma}[q_c^{ H_o}(\eta\eta')^F]~,
\ee
where $\mathcal{H}_i^\sigma$ is spanned by the conformal family corresponding to the `$i$' representation of the constraint algebra, and the $\chi_i$ are the characters.  As a point of clarification, note that in an Ishibashi state $i,~\eta$, and $\sigma$, denote the representation of a \emph{closed} string conformal family, the boundary condition on a \emph{closed} string state, and the \emph{closed} string sector (NS-NS or R-R), respectively.  In a Cardy state, $\eta$ and $\sigma$ have the same meaning while $\alpha$ labels an \emph{open} string conformal family.  This statement will become more clear in section \ref{subsec:ZZ and FZZT}.

These constraints imply that the Ishibashi states are constructed as \cite{Nakayama:2004vk}, \cite{Fukuda:2002bv},
\bea
\label{eq:N=1 Ishibashi}
|i;\eta,\textrm{NS} \rangle \rangle &=& |i;\textrm{NS}\rangle_L|i;\textrm{NS} \rangle_R + \mathrm{descendants} \nonumber \\ 
|i;\eta,\textrm{R} \rangle \rangle &=& a|i;\textrm{R}^+\rangle_L|i;\textrm{R}^+ \rangle_R - i\eta a|i;\textrm{R}^-\rangle_L|i;\textrm{R}^-\rangle_R \nonumber \\
& & + b|i;\textrm{R}^-\rangle_L|i;\textrm{R}^+\rangle_R - i\eta b|i;\textrm{R}^+\rangle_L|i;\textrm{R}^-\rangle_R + \mathrm{descendants}~,
\eea
where the coefficients $a$ and $b$ are determined by the constraint equations (note that this implies the coefficients of descendants in both sectors will have some $\eta$ dependence).  In fact, when we take the Type 0A projection we will have $a=0$ and when we take the Type 0B projection we will have $b=0$.  Now we may use these Ishibashi states to represent the Cardy states schematically as
\be
\label{eq:cardy}
|B,\alpha;\eta,\sigma\rangle = \sum_i \Psi_\alpha(i;\eta,\sigma)|i;\eta,\sigma\rangle\rangle~,
\ee
where $\alpha$ and $i$ may range over some combination of a continuous and discrete spectrum.  The trace in (\ref{eq:open-closed}) may also be represented as
\be
\label{eq:characters}
\textrm{Tr}_{\mathcal{H}_{\alpha\beta}^\sigma}[q_o^{H_o}(\pm1)^F] = \sum_i n^{i,\sigma}_{\alpha\beta} \chi_i^{\sigma,\pm}(q_o)~,
\ee
where the $n^{i,\sigma}_{\alpha\beta}$ are non-negative integers representing the multiplicty of $\mathcal{H}_i^\sigma$ in $\mathcal{H}_{\alpha\beta}^\sigma$.  Using (\ref{eq:open-closed}), (\ref{eq:cardy}), (\ref{eq:characters}), and the modular transformations of the open string characters, we may determine the `wave functions' $\Psi_\alpha(i;\eta,\sigma)$.  This is what is known as the modular bootstrap construction.

\subsection{ZZ and FZZT Boundary States}
\label{subsec:ZZ and FZZT}

As an example, let us demonstrate how the modular bootstrap is applied to determine the boundary states corresponding to the ZZ brane and the FZZT brane.  Since the stress tensor and the superconformal current of the $\hat{c}_m=1~\mathcal{N}=1$ SLFT are simply a sum of the corresponding currents of the $\hat{c}_m=1$ theory and the $\mathcal{N}=1$ SLFT, the tensor product of an Ishibashi state from each theory will be an Ishibashi state of the combined theory.  The ZZ and FZZT boundary states have been constructed for the $\mathcal{N}=1$ SLFT \cite{Ahn:2002ev}, \cite{Fukuda:2002bv}, and so a trivial modifaction allows us to write them in our theory.

The open string spectrum corresponding to the descendants of the vacuum state is given by an open string stretching between two vacuum Cardy states, while the spectrum corresponding to an excited state is given by an open string stretching between the corresponding excited Cardy state and a vacuum Cardy state:
\bea
\label{eq:N=1 O/C duality}
\chi_{\mathrm{vac}}^{\tilde{\sigma},\widetilde{\eta\eta'}}(\tau_o)&=&\langle
\mathrm{vac};\eta,\sigma|q_c^{\frac{1}{2}H_c}|\mathrm{vac};\eta',\sigma\rangle \nonumber \\
\chi_{p,\omega}^{\tilde{\sigma},\widetilde{\eta\eta'}}(\tau_o)&=&\langle\mathrm{vac};\eta,\sigma|q_c^{\frac{1}{2}H_c}|p,\omega;\eta',\sigma\rangle~,
\eea
where $\nu(\tilde{\sigma})=\frac{1}{2}|\eta-\eta'|$; $\nu(\sigma)=0,1$, corresponds to NS and R, respectively; and $\widetilde{\eta\eta'}=e^{i \pi \nu(\sigma)}$.

Our goal is to determine the wave function of the Cardy states expressed as a linear combination of the Ishibashi states, which we denote by
\be
|B,p,\omega;\eta,\sigma\rangle = \int_{-\infty}^{\infty}dp'~d\omega'~\Psi_{p,\omega}(p',\omega';\eta,\sigma)|p',\omega';\eta,\sigma\rangle\rangle~.
\ee
To apply this to the vacuum, note that the vacuum state has zero weight and so, in the NS sector, has momentum $p=-\frac{i}{2}(\frac{1}{b}+b),~\omega=0$.  Recall that this corresponds to a (1,1) degenerate representation since every conformal family built from a primary of momentum $p=-\frac{i}{2}(\frac{m}{b}+nb)$ is degenerate at level $mn$.  This means that the open string character corresponding to the vacuum representation is
\be
\label{eq:N=1 vac character}
\chi_{\textrm{vac}}^{\textrm{NS},+}(\tau) = [q^{-(\frac{1}{b}+b)^2/8}-q^{-(\frac{1}{b}-b)^2/8}]\frac{\theta_{00}(\tau,0)}{\eta(\tau)^3}~,
\ee
and the others are obtained similarly.  Then by inserting the expansion of the Cardy states into (\ref{eq:N=1 O/C duality}) and using the normalization of the Ishibashi states (\ref{eq:N=1 Ishibashi definition}), the above open/closed duality equations are rewritten as
\bea
\label{eq:N=1 s-transformation}
\chi_{\textrm{vac}}^{\tilde{\sigma},\widetilde{\eta\eta'}}(\tau_o) &=& \int_{-\infty}^\infty dp'~d\omega'~\Psi^{(\sigma)*}_{\textrm{vac}}(p',\omega';\eta)\Psi^{(\sigma)}_{\textrm{vac}}(p'\omega';\eta')\chi_{p',\omega'}^{\sigma,\eta\eta'}(\tau_c)  \nonumber \\
\chi_{p,\omega}^{\tilde{\sigma},\widetilde{\eta\eta'}}(\tau_o) &=& \int_{-\infty}^\infty dp'~d\omega'~\Psi^{(\sigma)*}_{\textrm{vac}}(p',\omega';\eta)\Psi^{(\sigma)}_{p,\omega}(p'\omega';\eta')\chi_{p',\omega'}^{\sigma,\eta\eta'}(\tau_c)~.
\eea

The modular transformations of the open string characters, given by the S-transformation matrix, determine the wave functions of the vacuum boundary states to be
\bea
\label{eq:N=1 vac wfn}
\Psi_{\textrm{vac}}^{\textrm{(NS)}}(p,\omega;\eta) &=& \frac{\pi (\mu\pi\gamma(\frac{bQ}{2}))^{-ip/b} }{ip\Gamma(-i p b)\Gamma(-i\frac{p}{b})}\Psi^{\textrm{(NS)},\hat{c}_m=1}_{\omega'=0}(\omega;\eta)  \nonumber \\
\Psi_{\textrm{vac}}^{\textrm{(R)}}(p,\omega;+) &=& \frac{\pi (\mu \pi\gamma(\frac{bQ}{2}))^{-ip/b}}{\Gamma(\frac{1}{2}-i p b) \Gamma(\frac{1}{2}-i\frac{p}{b})}\Psi^{\textrm{(R)},\hat{c}_m=1}_{\omega'=0}(\omega;+)~,
\eea
where $\Psi^{(\sigma),\hat{c}_m=1}_{\omega'}$ denotes the wave function for the $\hat{c}_m=1$ matter boundary state whose properties do not concern us here.  (Note that there is no R-sector $(1,1)$ boundary state with $\eta=-$ for the same reason that the open string character in this sector is zero \cite{Fukuda:2002bv}, \cite{Douglas:2003up}.)  Their pole structures show that these vacuum boundary states, the ZZ branes, correspond to static Euclidean D0-branes that sit in the strong coupling region, $\phi \rightarrow +\infty$.

All the excited states are non-degenerate representations with
continuous momentum $p'$.  The modular transformation of the open string
character of the continuous representation together with the
solution of the ZZ boundary state gives the excited boundary state
(FZZT brane)
\bea
\label{eq:N=1 FZZT wfn}
\Psi_{p',\omega'}^{\textrm{(NS)}}(p,\omega;\eta) & = & -\frac{\cos(2\pi p p')}{2\pi} \textstyle\left(\mu\pi\gamma\left(\frac{bQ}{2}\right)\right)^{-ip/b} i p \Gamma(i p b) \Gamma\left(i\frac{p}{b}\right) \displaystyle\Psi^{\textrm{(NS)},\hat{c}_m=1}_{\omega'}(\omega;\eta)  \nonumber \\
\Psi_{p',\omega'}^{\textrm{(R)}}(p,\omega;+) & = & \frac{\cos(2\pi p p')}{2\pi}\textstyle \left(\mu\pi\gamma\left(\frac{bQ}{2}\right)\right)^{-ip/b}\Gamma\left(\frac{1}{2}+i p b\right) \Gamma\left(\frac{1}{2}+i\frac{p}{b}\right) \displaystyle\Psi^{\textrm{(R)},\hat{c}_m=1}_{\omega'}(\omega;+)~.
\eea
Their pole structures show that they are Euclidean D1-branes, extended in the Liouville direction.

\subsection{An Argument for Additional Symmetry}
\label{subsec:Existence}

As we saw above, the characters for $\hat{c}_m=1~\mathcal{N}=1$ SLFT are simply a product of the individual characters for $\hat{c}_m=1$ matter and $\mathcal{N}=1$ SLFT.  This led us to a trivial modification of the ZZ and FZZT boundary states of $\mathcal{N}=1$ SLFT that resulted in static branes, as we saw in section \ref{subsec:ZZ and FZZT}.  In fact, we were destined to realize this result because we restricted ourselves to the subset of $\hat{c}_m=1~\mathcal{N}=1$ Ishibashi states that were simply a tensor product of the Ishibashi states of the separate theories.  We thus implicitly required our boundary states to satisfy an additional symmetry (namely, that they separately satisfy $\mathcal{N}=1$ boundary conditions for each direction).  It was because of this additional symmetry we imposed, combined with the fact the characters decouple, that the wave functions did not mix the two directions.

If we want to find a falling D0-brane, we clearly cannot impose the restrictions mentioned above.  Naturally, the time direction should satisfy `Neumann-like' boundary conditions while the Liouville direction should satisfy `Dirichlet-like' conditions.  However, what conditions to impose are not obvious.  In the bosonic case of the hairpin brane \cite{Lukyanov:2003nj}, the authors had to impose the $\mathcal{W}$-symmetry to get the time dependence necessary to obtain a falling D0-brane with the desired trajectory.

It turns out that in the $\mathcal{N}=2$ SLFT, which has the same matter content as $\hat{c}_m=1~\mathcal{N}=1$ SLFT, there are additional symmetries that naturally follow from the action.  Applying these additional constraints to $\mathcal{N}=1$ Ishibashi states, one finds boundary states with trajectories that match that of the falling D0-brane (\ref{eq:classical trajectory}).  We will show that in $\mathcal{N}=1$, 2D superstring theory with linear dilaton background (which is equivalent to $\hat{c}_m=1~\mathcal{N}=1$ SLFT), there exists a type of falling D0-brane that has the additional $\mathcal{N}=2$ SCA symmetry and can be obtained by a slight modification of the falling D0-brane solution of the $\mathcal{N}=2$ SLFT.

\section{$\mathcal{N}=2$ SLFT and its Boundary States}
\label{sec:N=2}

\subsection{$\mathcal{N}=2$ SLFT}

The $\mathcal{N}=2$ SLFT theory has the same free action as $\hat{c}_m=1~\mathcal{N}=1$ SLFT given by (\ref{eq:N=1 action}), but has different interaction terms.  In fact, there are two types of interaction terms that are consistent with the $\mathcal{N}=2$ superconformal symmetry.  The chiral interaction terms are
\bea
\label{eq:N=2 interaction chiral}
S_c^{\mathcal{N}=2} &=& 2\mu b^2 \int d^2z \Big( \frac{\pi}{2}\mu :e^{b(\phi+iY)}::e^{b(\phi-iY)}: + (\psi^1\tilde{\psi}^1-\psi^2\tilde{\psi}^2)e^{b\phi}\cos bY \nonumber \\
& & - (\psi^2\tilde{\psi}^1+\psi^1\tilde{\psi}^2)e^{b\phi}\sin bY \Big)~,
\eea
while the non-chiral interaction terms are
\be
\label{eq:N=2 interaction nonchiral}
S_{nc}^{\mathcal{N}=2} = \mu' \int d^2z\left(\partial\phi - i\partial Y + \frac{i}{b}\psi^1\psi^2\right)\left(\bar{\partial}\phi + i\bar{\partial}Y + \frac{i}{b}\tilde{\psi}^1\tilde{\psi}^2\right)e^{\frac{1}{b}\phi}~.
\ee
In this theory, the background charge does not get renormalized, so we have $Q=\frac{1}{b}$ instead of $Q=\frac{1}{b}+b$ as we had in the $\mathcal{N}=1$ SLFT and the bosonic LFT.  Only the non-chiral interaction preserves the $\mathcal{N}=2$ supersymmetry after a Wick rotation of the Euclidean time, $Y$.

Since the free part of the action is the same as for $\hat{c}_m=1~\mathcal{N}=1$ SLFT, $T$ and $G$---which we will call $G^1$ for this section---are the same as before.  However, since this is an $\mathcal{N}=2$ theory, we have a current $G^2$ corresponding to the second supercharge.  Both the chiral and non-chiral interaction terms are invariant under a combined shift in Euclidean time and rotation between the two fermions, leaving us with an additional $U(1)$ current, $J$:
\bea
T & = & -\frac{1}{2}\partial\phi\partial\phi - \frac{1}{2}\partial Y \partial Y - \frac{1}{2}\delta_{\mu\nu}\psi^\mu\partial\psi^\nu + \frac{Q}{2}\partial^2\phi \nonumber \\
G^1 & = & i\left(\psi^1\partial\phi + \psi^2\partial Y - Q\partial\psi^1\right) \nonumber \\
G^2 & = & -i\left(\psi^2\partial\phi - \psi^1\partial Y - Q\partial\psi^2\right) \nonumber \\
J & = & i\left(\psi^1\psi^2 + Q\partial Y\right)~.
\eea
It will be convenient to define $G^\pm \equiv \frac{1}{\sqrt{2}}(G^1\pm iG^2)$, which allows us to write the $\mathcal{N}=2$ SCA as
\bea
\label{eq:N=2 SCA algebra}
[ L_m,L_n ] & = & (m-n)L_{m+n} + \frac{c}{12}(m^3-m)\delta_{m,-n}~, \nonumber \\
\lbrack L_m,G_r^\pm \rbrack & = & \left(\frac{m}{2}-r\right) G_{m+r}^\pm ~, \qquad \lbrack J_m,G_r^\pm \rbrack = \pm G_{m+r}^\pm~,  \nonumber \\
\{ G_r^+,G_s^- \} & = & 2L_{r+s} + (r-s)J_{r+s} + \frac{c}{12}(4r^2-1)\delta_{r,-s}~,  \qquad  \{ G_r^\pm,G_s^\pm\}=0~, \nonumber \\
\lbrack L_m,J_n\rbrack & = & -nJ_{m+n}~,  \qquad  [J_m,J_n]=\frac{c}{3}m\delta_{m,-n}~,
\eea
with central charge $c=3\hat{c}=3(1+Q^2)$, so again $Q=2$ corresponds to a critical string theory, but in this case we must have $b=\frac{1}{2}$.  The primary fields are the same as in section \ref{subsec:c=1 N=1 SLFT}, with corresponding $U(1)$ charges
\bea
\label{eq:N=2 charges}
j_{p,\omega}^{\textrm{NS}} &=& Q \omega  \nonumber \\
j_{p,\omega}^{\textrm{R},\pm} &=& j^{\textrm{NS}}_{p,\omega}\pm\frac{1}{2}~.
\eea

The open string character is
\be
\label{eq:N=2 trace}
\chi_\xi^{\sigma,\pm}(\tau,\nu)=\mathrm{Tr}_{\mathcal{H}_\xi^\sigma}[q^{L_0-c/24}y^{J_0}(\pm1)^F]~,
\ee
where $q=e^{2 \pi i \tau}$, $y=e^{2 \pi i \nu}$, and $\xi$ denotes the `$\xi$' representation of the constraint algebra.  The characters of $\mathcal{N}=2$ SCA representations split into three classes using the fermionic operator $G_{-\frac{1}{2}}^\pm$ \cite{Eguchi:2003ik}, \cite{Nakayama:2004vk}:

\begin{itemize}

\item \textbf{Class 1 (Graviton):} The graviton representation is defined by the open string primaries satisfying $G_{-\frac{1}{2}}^{\pm} |\textrm{graviton}\rangle = 0$, which implies that $L_{-1} |\textrm{graviton}\rangle = 0$.  This constrains the momentum to $p=-\frac{i}{2b},~ \omega=0$, which implies $h=0$. Thus, the graviton representation corresponds to the unique vacuum state (the identity operator).  After eliminating this state, the character becomes
\be
\label{eq:N=2 III character}
\chi_\textrm{vac}^{\textrm{NS},+}(\tau,\nu) =  q^{-\frac{1}{8b^2}} \frac{1-q}{(1+y\sqrt{q})(1+y^{-1}\sqrt{q})}\frac{\theta_{00}(\tau,\nu)}{\eta(\tau)^3}~.
\ee
(Note that in all three classes, $\chi_\xi^{\textrm{NS},-}$ is obtained by replacing $\theta_{00}$ with $\theta_{01}$, $\chi_\xi^{\textrm{R},+}$ by replacing both $\theta_{00}$ with $\theta_{10}$ and $j=Q\omega$ with $j=Q\omega\pm\frac{1}{2}$ (chiral/anti-chiral), and $\chi_\xi^{\textrm{R},-}=0$.)

\item \textbf{Class 2 (Massive):} The massive representation is defined by $G_{-\frac{1}{2}}^{\pm} |\textrm{massive}\rangle \neq 0$.  For generic $p$ and $\omega$, this representation is non-degenerate.  The NS character is obtained by summing over all the descendants while using $j=Q\omega$:
\be
\label{eq:N=2 I character}
\chi_{[p,\omega]}^{\textrm{NS},+}(\tau,\nu) = q^{\frac{1}{2}(p^2+\omega^2)}y^{Q\omega}\frac{\theta_{00}(\tau,\nu)}{\eta(\tau)^3}~.
\ee

\item \textbf{Class 3 (Massless):} The massless representation is defined by $G_{-\frac{1}{2}}^+|\textrm{chiral}\rangle = 0$ or $G_{-\frac{1}{2}}^-|\textrm{anti-chiral}\rangle = 0$ for the chiral or anti-chiral representations, respectively. This implies that the momentum must satisfy $\frac{Q}{2}+ip=\pm \omega$, respectively.  The character for the chiral representation is obtained by eliminating the contribution from the $G_{-\frac{1}{2}}^{+}$ mode:
\be
\label{eq:N=2 II character}\chi_{\omega}^{\textrm{NS},+}(\tau,\nu) =  q^{-\frac{1}{8b^2}} \frac{(y\sqrt{q})^{Q\omega}}{1+y\sqrt{q}}\frac{\theta_{00}(\tau,\nu)}{\eta(\tau)^3}~.
\ee

\end{itemize}

\subsection{$\mathcal{N}=2$ Ishibashi States and Cardy States}
\label{subsec:N=2 I/C}

An $\mathcal{N}=2$ boundary state is constructed as in the $\mathcal{N}=1$ system.  Naturally, an $\mathcal{N}=2$ boundary state must satisfy the $\mathcal{N}=1$ conditions
\be
\label{eq:N=1 b.s. 1}
(L_m - \tilde{L}_{-m}) |B;\eta,\sigma\rangle = (G_r^1-i\eta\tilde{G}_{-r}^1)|B;\eta,\sigma\rangle = 0~.
\ee
Additionally, an $\mathcal{N}=2$ boundary state will satisfy one of two different conditions.  An A-Type boundary state will satisfy
\be
\label{eq:N=2 b.s. A}
(J_m - \tilde{J}_{-m})|B;\eta,\sigma\rangle = (G^\pm_r - i\eta\tilde{G}^\mp_{-r})|B;\eta,\sigma\rangle = 0~,
\ee
while a B-Type state will satisfy
\be
\label{eq:N=2 b.s. B}
(J_m + \tilde{J}_{-m})|B;\eta,\sigma\rangle = (G^\pm_r - i\eta\tilde{G}^\pm_{-r})|B;\eta,\sigma\rangle = 0~.
\ee
B-Type conditions correspond to `Neumann-like' boundary conditions on Euclidean time while A-Type conditions correspond to `Dirichlet-like' boundary conditions.  Since we are interested in studying D0-branes, we will focus on the B-Type states for the rest of this paper.

If we denote the R-sector primary states as $|h,j;R^\pm\rangle_L$, where $j=Q\omega$ as in (\ref{eq:N=2 charges}) and $\pm$ denotes the spin structure, then $J_0 |h,j;R^\pm\rangle_L = (j\pm\frac{1}{2})|h,j;R^\pm\rangle_L$ (and similarly in the right-moving sector).  We can check from (\ref{eq:N=1 Ishibashi}) that the B-Type, R-R sector Ishibashi states can be constructed schematically as
\be
\label{eq:N=2 B Ishibashi}
|h,j;\eta,\textrm{R} \rangle \rangle \propto |h,j;\textrm{R}^-\rangle_L|h,-j;\textrm{R}^+\rangle_R - i\eta |h,j;\textrm{R}^+\rangle_L|h,-j;\textrm{R}^-\rangle_R + \mathrm{descendants}~,
\ee
Since $\psi^+_0|h,j;\textrm{R}^-\rangle_L = |h,j;\textrm{R}^+\rangle_L$, it is clear that $(-1)^{F+\tilde{F}}=-1$ on the primary states in (\ref{eq:N=2 B Ishibashi}).  Furthermore, from the commutation relations $\{J_0,G_0^\pm\}=\pm G_0^\pm$, we can see that the constraint $(J_0+\tilde{J}_0)=0$ implies that all descendants in (\ref{eq:N=2 B Ishibashi}) must have an equal number of fermionic raising operators on the left-moving and right-moving sides, modulo 2.  Therefore, B-Type, R-R sector Ishibashi states will be projected out by the Type 0B GSO projection and so are only present in Type 0A---note that a similar argument implies that A-Type, R-R sector Ishibashi states are only present in Type 0B.  Thus, B-Type states will yield stable D0-branes in Type 0A and unstable D0-branes in Type 0B.

B-Type Ishibashi states are constructed to form an orthonormal basis for states satisfying (\ref{eq:N=1 b.s. 1}) and (\ref{eq:N=2 b.s. B}), and must also satisfy
\bea
\label{eq:N=2 Ishibashi}
& \textbf{Class 1} & \quad \langle \langle\textrm{vac};\eta,\sigma|q_c^{\frac{1}{2}H_c} y_c^{\frac{1}{2}(J_0-\tilde{J}_0)} |\textrm{vac};\eta,\sigma\rangle
\rangle = \chi_{\textrm{vac}}^{\sigma,\eta\eta'}(\tau_c,\nu_c)  \nonumber \\
& \textbf{Class 2} & \quad \langle \langle p,\omega;\eta,\sigma|q_c^{\frac{1}{2}H_c} y_c^{\frac{1}{2}(J_0-\tilde{J}_0)} |p',\omega';\eta,\sigma\rangle \rangle = \delta(p-p')\delta(\omega-\omega')\chi_{[p,\omega]}^{\sigma,\eta\eta'}(\tau_c,\nu_c) \nonumber \\
& \textbf{Class 3} & \quad \langle \langle\omega;\eta,\sigma|q_c^{\frac{1}{2}H_c}y_c^{\frac{1}{2}(J_0-\tilde{J}_0)} |\omega';\eta,\sigma\rangle \rangle = \delta(\omega-\omega')\chi_\omega^{\sigma,\eta\eta'}(\tau_c,\nu_c)~,
\eea
while all other correlators between Ishibashi states vanish.  The open and closed parameters are related by the modular transformation $\tau_o=-\frac{1}{\tau_c}$ and $\nu_o=\frac{\nu_c}{\tau_c}$.  The B-Type Cardy states are then constructed as a linear combination of the B-Type Ishibashi states such that the Cardy states satisfy
\bea
\langle B,O;\eta,\sigma|q_c^{\frac{1}{2}H_c}y_c^{\frac{1}{2}(J_0-\tilde{J}_0)}|B,\xi;\eta',\sigma\rangle &=& e^{i\pi\tilde{c}\frac{\nu_o^2}{\tau_o}}\chi_\xi^{\tilde{\sigma},\widetilde{\eta\eta'}}(\tau_o,\nu_o) \nonumber \\
\langle B,O;\eta,\sigma|q_c^{\frac{1}{2}H_c}y_c^{\frac{1}{2}(J_0-\tilde{J}_0)}|B,O;\eta',\sigma\rangle &=& e^{i\pi\tilde{c}\frac{\nu_o^2}{\tau_o}}\chi_\textrm{vac}^{\tilde{\sigma},\widetilde{\eta\eta'}}(\tau_o,\nu_o)~,
\eea
where $\chi_\xi^{\sigma,\pm}(\tau,\nu)$ is the open string character of the $\xi$ representation of the constraint algebra, $O$ represents the graviton state, and $\tilde{\sigma}$ and $\widetilde{\eta\eta'}$ are defined as in equation (\ref{eq:N=1 O/C duality}).

\subsection{Falling Euclidean D0-Brane in $\mathcal{N}=2$ SLFT}
\label{subsec:ED0-Brane}

As in $\hat{c}_m=1~\mathcal{N}=1$ SLFT, the modular transformation of
the Class 1 (Graviton) representation (identity operator) gives
the wave function of the vacuum boundary state. Then the modular
transformation of the Class 2 (Massive) non-degenerate
representation of the open string produces the wave function of the
excited boundary state which corresponds to the FZZT brane
(Falling Euclidean D0-brane) solution \cite{Eguchi:2003ik}, \cite{Nakayama:2004yx}, \cite{Ahn:2003tt}, \cite{Ahn:2004qb}:
\be
\label{eq:N=2 FZZT wfn}
\Psi_{[p',\omega']}(p,\omega;\eta,\sigma)  =  \sqrt{2}Q e^{-2\pi i\omega\omega' }\cos(2\pi p p') \frac{\Gamma(-iQp)\Gamma\left(1-i\frac{2p}{Q}\right)}{\Gamma\left(\frac{1}{2}-i\frac{p}{Q}+\frac{\omega}{Q}-\frac{\nu(\sigma)}{2}\right)\Gamma\left(\frac{1}{2}-i\frac{p}{Q}-\frac{\omega}{Q}+\frac{\nu(\sigma)}{2}\right)}~.
\ee
Note that this wave function has no dependence on $\eta$.

The Fourier transform of the momentum space wave function into the position
space wave function is
\be
\label{eq:N=2 FZZT Fourier}
\tilde{\Psi}_{[p',\omega']}^{(\textrm{NS})}(\phi,Y)\equiv\int_{-\infty}^{\infty} \frac{dpd\omega}{(2\pi)^2} \,e^{-ip\phi}e^{-i\omega Y}\Psi_{[p',\omega']}^{(\textrm{NS})}(p,\omega)~.
\ee
We can construct solutions where $p'$ and $\omega'$ are nonzero from the solution in which they are both zero \cite{Nakayama:2004yx}
\be
\label{eq:N=2 FZZT wfn position}
\tilde {\Psi}^{(\textrm{NS})}_{[0,0]}(\phi,Y)=\frac{\sqrt{2}} {\pi Q (2\cos\frac{Q Y}{2})^{\frac{2}{Q^2}+1}}\cdot\exp\left[-\frac{\phi}{Q}-\frac{e^{-\frac{\phi}{Q}}} {(2\cos\frac{QY}{2})^{\frac{2}{Q^2}}}\right]~.
\ee
Then it is simple to see that
\be
\tilde{\Psi}^{(\textrm{NS})}_{[p',\omega']}(\phi,Y) = \frac{1}{2}\tilde{\Psi}^{(\textrm{NS})}_{[0,0]}(\phi-2\pi p',Y+2\pi\omega') + \frac{1}{2}\tilde{\Psi}^{(\textrm{NS})}_{[0,0]}(\phi+2\pi p',Y+2\pi\omega')~.
\ee
The classical shape of the falling Euclidean D0-brane corresponding to $[p',\omega']=[0,0]$ is given by the peak of the position space wave function (\ref{eq:N=2 FZZT Fourier}):
\be
\label{eq:D1-brane shape}
e^{-\frac{Q \phi}{2}}=2\cos \frac{Q Y}{2}~.
\ee

\subsection{Falling D0-brane in $\mathcal{N}=2$ SLFT}
\label{subsec:D0-Brane}

The Wick-rotation from the Euclidean time, $Y$, into the Minkowski time,
$t$, together with a shift in the Liouville direction, $\phi \to \phi-\frac{2}{Q}\ln\tilde{r}$, produces the classical trajectory of the falling D0-brane
in $\mathcal{N}=2$ SLFT \cite{Nakayama:2004yx}:
\be
\label{eq:D0-brane trajectory}
\tilde{r}e^{-\frac{Q \phi}{2}}=2\cosh \frac{Q t}{2}~,
\ee
which matches with (\ref{eq:classical trajectory}) once we set $\tilde{r}=\frac{2E}{\tau_p}$.

Therefore, the falling D0-brane wave function in position space is \cite{Nakayama:2004yx}
\be
\label{eq:falling D0-brane wfn position}
\tilde {\Psi}^{(\textrm{NS})}_{[0,0]}(\phi,t) = \frac{\sqrt{2}}{\pi Q (2\cosh\frac{Q t}{2})^{\frac{2}{Q^2}+1}} \cdot\exp\left[-\frac{\phi-\frac{2}{Q}\ln\tilde{r}}{Q}-\frac{e^{-\frac{\phi-\frac{2}{Q}\ln\tilde{r}}{Q}}}{(2\cosh \frac{Qt}{2})^{\frac{2}{Q^2}}}\right]~.
\ee
Then the Fourier transform to momentum space yields
\be
\label{eq:falling wfn NS}
\Psi^{(\textrm{NS})}_{[0,0]}(p,q)=\frac{-i \sqrt{2} Q e^{i\frac{2 p}{Q}\ln\tilde {r} }\sinh(\frac{ 2 \pi p}{Q})} {\cosh(\frac{2\pi p}{Q}) + \cosh(\frac{2\pi q}{Q})} \cdot \frac{\Gamma(-iQp)\Gamma(1-i\frac{2p}{Q})}{\Gamma(\frac{1}{2}-i\frac{p}{Q}+i\frac{q}{Q})\Gamma(\frac{1}{2}-i\frac{p}{Q}-i\frac{q}{Q})}~,
\ee
and a half spectral flow gives the R-sector wave function
\be
\label{eq:falling wfn R}
\Psi^{(\textrm{R})}_{[0,0]}(p,q) = \frac{-i \sqrt{2} Q e^{i\frac{2 p}{Q}\ln\tilde{r} }\sinh (\frac{2\pi p}{Q})} {\cosh(\frac{2\pi p}{Q}) - \cosh(\frac{2\pi q}{Q})} \cdot \frac{\Gamma(-iQp)\Gamma(1-i\frac{2p}{Q})} {\Gamma(1-i\frac{p}{Q}+i\frac{q}{Q})\Gamma(-i\frac{p}{Q}-i\frac{q}{Q})}~.
\ee

\section{Falling D0-brane in $\mathcal{N}=1$, 2D Superstring Theory}
\label{sec:Falling D0-Brane}

\subsection{Using $\mathcal{N}=2$ SLFT to Study Boundary States in 2D Superstring Theory}
\label{subsec:N=1 from N=2}

We propose that the $\mathcal{N}=2$ SLFT boundary states may be used to study falling D0-branes in the $\mathcal{N}=1$, 2D superstring with linear dilaton background (which is equivalent to $\hat{c}_m=1~\mathcal{N}=1$ SLFT theory).  Notice that the field content of both theories is the same, as are the stress tensor and the first supercharge.  Additionally, as is apparent from the constraints on an $\mathcal{N}=2$ boundary state, any $\mathcal{N}=2$ boundary state also satisfies the $\mathcal{N}=1$ constraints (\ref{eq:N=1 b.s.}).  So an $\mathcal{N}=2$ SLFT boundary state is also a boundary state of the $\mathcal{N}=1$, 2D superstring.

However, this alone is not enough; if we want to use the $\mathcal{N}=2$ Ishibashi states, we must also be able to construct a Cardy state from them that will generate the $\hat{c}_m=1~\mathcal{N}=1$ open string character.  In fact, we can do this.  In (\ref{eq:N=2 I character}), $\nu$ is the `modulus' of the $U(1)$ charge and appears nontrivially in the functional form of the character.  But if we set $\nu=0$ ($y=1$), the $U(1)$ charge acts trivially on the $\mathcal{N}=2$ states and we can see that the character of the $\mathcal{N}=2$ Class 2 (Massive) representation (\ref{eq:N=2 I character}) is equivalent to the $\hat{c}_m=1~\mathcal{N}=1$ character (\ref{eq:N=1 character}).  Such a state was found in \cite{Eguchi:2003ik}, \cite{Nakayama:2004yx}, \cite{Ahn:2003tt}, \cite{Ahn:2004qb}, and presented in sections \ref{subsec:ED0-Brane} and \ref{subsec:D0-Brane}.  The momentum space wave functions are the same as in (\ref{eq:falling wfn NS}) and (\ref{eq:falling wfn R}):
\bea
\Psi^{(\textrm{NS})}_{[0,0]}(p,q) & = & \frac{-i \sqrt{2} Q e^{i\frac{2 p}{Q}\ln\tilde {r} }\sinh(\frac{ 2 \pi p}{Q})} {\cosh(\frac{2\pi p}{Q}) + \cosh(\frac{2\pi q}{Q})} \cdot \frac{\Gamma(-iQp)\Gamma(1-i\frac{2p}{Q})}{\Gamma(\frac{1}{2}-i\frac{p}{Q}+i\frac{q}{Q})\Gamma(\frac{1}{2}-i\frac{p}{Q}-i\frac{q}{Q})}   \nonumber \\
\Psi^{(\textrm{R})}_{[0,0]}(p,q) & = & \frac{-i \sqrt{2} Q e^{i\frac{2 p}{Q}\ln\tilde{r} }\sinh (\frac{2\pi p}{Q})} {\cosh(\frac{2\pi p}{Q}) - \cosh(\frac{2\pi q}{Q})} \cdot \frac{\Gamma(-iQp)\Gamma(1-i\frac{2p}{Q})} {\Gamma(1-i\frac{p}{Q}+i\frac{q}{Q})\Gamma(-i\frac{p}{Q}-i\frac{q}{Q})}~.
\eea

This is not a surprising result.  Recall that in section \ref{subsec:Existence}, we argued that in $\hat{c}_m=1~\mathcal{N}=1$ SLFT we could find a falling boundary state with the Liouville and time directions coupled nontrivially by assuming additional symmetries that coupled the two directions.  This additional symmetry is a symmetry only of the boundary state and not of the theory as a whole.  The coupling of the Liouville and time directions is then achieved by imposing this additional symmetry on the Hilbert space of the original $\hat{c}_m=1~\mathcal{N}=1$ SLFT boundary states.

Note that it is also possible to derive this falling D0-brane in the $\mathcal{N}=1$, 2D superstring by directly solving the constraint equations satisfied by the boundary states, similar to the derivation of the bosonic hairpin brane \cite{Lukyanov:2003nj}.  The equations are constrained by the $\mathcal{W}$-symmetry in the bosonic case, while by the $\mathcal{N}=2$ SCA in the $\mathcal{N}=1$, 2D superstring.

\subsection{Number of D0-branes after GSO projection}

In $\mathcal{N}=1$, 2D superstring theory, there are two distinct types of boundary states in each of the NS-NS and R-R sectors, corresponding to the different boundary conditions for world sheet fermions ($\eta=\pm$).  Therefore, the Type 0, non-chiral GSO projection produces four types of stable D0-branes (two branes and two anti-branes) in the Type 0A theory, and two unstable D0-branes in the Type 0B theory.  In the Type 0A theory, the D$0^{\pm}$-branes are sourced by two different R-R gauge fields, $C_{1}^{(\pm)}$ \cite{Thompson:2001rw}.

The D-brane boundary states in the Type 0A theories are given by the non-chiral GSO projection $\frac{1 \pm (-1)^{F+\tilde{F}}}{2}$, with the upper sign for the NS-NS sector and the lower sign for the R-R sector.  As explained in section \ref{subsec:N=2 I/C}, the D0-branes of our theory correspond to the B-Type boundary states.  Since the B-Type, R-R Ishibashi states survive the Type 0A GSO projection, the D0-branes in Type 0A will be stable.  We can represent them schematically as \cite{Gaberdiel:2000jr}, \cite{Thompson:2001rw},
\bea
\label{eq:0A b.s.}
|D0;+\rangle &=& |B0;+,\textrm{NS}\rangle + |B0;+,\textrm{R}\rangle\nonumber \\ 
|D0;-\rangle &=& |B0;-,\textrm{NS}\rangle + |B0;-,\textrm{R}\rangle \nonumber \\
|\overline{D}0;+\rangle &=& |B0;+,\textrm{NS}\rangle - |B0;+,\textrm{R}\rangle \nonumber \\
|\overline{D}0;-\rangle &=& |B0;-,\textrm{NS}\rangle - |B0;-,\textrm{R}\rangle~.
\eea

On the other hand, the D-brane boundary states in the Type 0B theories are given by the non-chiral GSO projection $\frac{1+(-1)^{F+\tilde{F}}}{2}$ for both the NS-NS and R-R sectors.  In this case, the B-Type, R-R Ishibashi states are projected out by the Type 0B projection.  This leaves us with two unstable D0-branes represented schematically as
\bea
\label{eq:0B b.s.}
|\widehat{D0};+\rangle &=& |B0;+,\textrm{NS}\rangle \nonumber \\
|\widehat{D0};-\rangle &=& |B0;-,\textrm{NS}\rangle~.
\eea
Note that the sign, $\eta=\pm$, really does denote different D0-branes since, by Cardy's condition (\ref{eq:open-closed}) and (\ref{eq:N=1 O/C duality}), these states yield different spectra.  It would be interesting to see how states with different values of $\eta$ distinguish themselves from each other in the context of matrix models.

\section{Discussion and Summary}

Recall that in the static case, the D0-brane (ZZ brane) and the D1-brane (FZZT brane) boundary states were derived from the degenerate and the non-degenerate representation of the open string character, respectively.  So it may seem a little puzzling that the falling D0-brane is constructed from the non-degenerate representation instead of the degenerate one.

In the static case, the difference between the D0-brane and the D1-brane is that the D0-brane is localized at the same point in the Liouville direction for all time, while the D1-brane is extended.  The open strings ending on this D0-brane can only take on a fixed (imaginary) value of Liouville momentum, while the open strings ending on the D1-brane can take on any value of Liouville momentum.

In the case of the falling D0-brane, if we partition the two-dimensional spacetime into spacelike hypersurfaces, the falling D0-brane is again localized in the Liouville direction along each hypersurface.  However, the Liouville position from one hypersurface to the next is not the same (the falling D0-brane is, not surprisingly, \emph{moving}), and so open strings ending on the falling D0-brane can take on any value of Liouville momentum.  This is the reason that we must use the non-degenerate representation of the open string characters to define the falling D0-brane boundary state.

To summarize, we have shown that a falling D0-brane boundary state in $\mathcal{N}=1$, 2D superstring theory can be obtained by adapting the falling D0-brane boundary state solution in $\mathcal{N}=2$ SLFT \cite{Nakayama:2004yx}.  In particular, there exist four types of stable, falling D0-branes (two branes and two anti-branes) in Type 0A theory and two types of unstable, falling D0-branes in Type 0B theory.  As is well known, Type 0, $\mathcal{N}=1$, 2D superstring theory has a dual description in the language of matrix models.  An interesting question would be to understand these falling D0-branes in the context of the dual matrix model.

\section{Acknowledgments}

We would like to thank A. Strominger and T. Takayanagi for many enlightening discussions.  The work of JML is supported by an NDSEG Fellowship sponsored by the Department of Defense.

\bibliography{paperrefs.v1}

\end{document}